\documentclass[twocolumn,showpacs,floatfix,prl,superscriptaddress]{revtex4}

\usepackage{float}
\usepackage{graphicx}

\newcommand{\bk}{{\bf k}}

\newcommand{\beq}{\begin{eqnarray}}
\newcommand{\eeq}{\end{eqnarray}}
\newcommand{\beqq}{\begin{eqnarray*}}
\newcommand{\eeqq}{\end{eqnarray*}}

\begin{document}

\title{Majorana Fermions and Exotic Surface Andreev Bound States in Topological Superconductors: Application to Cu$_x$Bi$_2$Se$_3$}

\author{Timothy H. Hsieh}
\affiliation{Department of Physics, Massachusetts Institute of Technology, Cambridge, MA 02139}

\author{Liang Fu}
\affiliation{Department of Physics, Massachusetts Institute of Technology, Cambridge, MA 02139}
\affiliation{Department of Physics, Harvard University, Cambridge, MA 02138}


\begin{abstract}

The recently discovered superconductor Cu$_x$Bi$_2$Se$_3$ is a candidate for three-dimensional time-reversal-invariant topological superconductors, which are predicted to have robust surface Andreev bound states hosting massless Majorana fermions.  In this work, we analytically and numerically find the linearly dispersing Majorana fermions at $k=0$, which smoothly evolve into a new branch of gapless surface Andreev bound states near the Fermi momentum. The latter is a new type of Andreev bound states resulting from both the nontrivial band structure and the odd-parity pairing symmetry.  The tunneling spectra of these surface Andreev bound states agree well with a recent point-contact spectroscopy experiment\cite{ando} and yield additional predictions for low temperature tunneling and photoemission experiments. 

\end{abstract}

\maketitle

The discovery of topological insulators 
 has generated much interest in not only understanding their properties and potential applications to spintronics and thermoelectrics but also 
searching for new topological phases.  A particularly exciting avenue is topological superconductors\cite{ludwig, kitaev, read, roy, zhangtsc, volovik, yip, moore, nagaosa}, 
in which unconventional pairing symmetries lead to topologically ordered superconducting ground states\cite{fuberg, sato, qihugheszhang}. 
The hallmark of a topological superconductor is the existence of gapless surface Andreev bound states which host itinerant Bogoliubov quasiparticles. 
These quasiparticles are solid-state realizations of massless Majorana fermions.  

There is currently an intensive search for topological superconductors.  
In particular, a recently discovered superconductor Cu$_x$Bi$_2$Se$_3$ with $T_c \sim 3K$\cite{cava} has attracted much attention\cite{palee}.  
A theoretical study\cite{fuberg} proposed that the strong spin-orbit coupled band structure of Cu$_x$Bi$_2$Se$_3$ favors  an 
odd-parity pairing symmetry, which leads to a time-reversal-invariant topological superconductor in three dimensions. 
Subsequently, many experimental and 
theoretical efforts\cite{wray, heat, sample, magnetization, haolee} have been made towards understanding superconductivity in Cu$_x$Bi$_2$Se$_3$. 
In a very recent point-contact spectroscopy experiment, Sasaki {\it et al.}\cite{ando} have 
observed a zero-bias conductance peak which strongly indicates unconventional pairing\cite{abs}.  
 
In this Letter, we find a new branch of gapless surface Andreev bound states (SABS), in addition to linearly dispersing Majorana fermions at $\bk=0$, in the topological superconducting phase of Cu$_x$Bi$_2$Se$_3$ and related doped semiconductors. This new branch of SABS is located near the Fermi momentum and is protected by a new bulk topological invariant.  Moreover, they result in unique features in the tunneling spectra which are in good agreement with the point-contact spectroscopy experiment on Cu$_x$Bi$_2$Se$_3$\cite{ando}.  We conclude by predicting clear signatures of these SABS, which can be tested in future tunneling and photoemission experiments at low temperatures. 
 

We start from the $k\cdot p$ Hamiltonian for the band structure of Cu$_x$Bi$_2$Se$_3$ near $\Gamma$\cite{fuberg}
\beq
H(\bk)= m \sigma_x + v_z k_z \sigma_y + v \sigma_z (k_x s_y - k_y s_x). \label{kp}
\eeq
Here $\sigma_z=\pm1$ labels the two Wannier functions which are primarily $p_z$ orbitals (from Se and Bi atoms) on 
the upper and lower part of the quintuple layer (QL) unit cell respectively (see Fig.1). Each orbital has a two-fold spin degeneracy labeled by $s_z=\pm 1$.  
We note that an earlier $k\cdot p$ Hamiltonian\cite{zhang} violates the mirror symmetry of the lattice, and a corrected version\cite{liu} is consistent with (\ref{kp}).  
Detailed discussion of the discrepancy is left to Supplementary Material\cite{detail}. 
The sign of  $mv_z$ is a crucial quantity which will now be inferred from   
the existence of surface states near $k_x=k_y=0$ in the surface Brillouin zone.  


\begin{figure}
\centering
\includegraphics[width=3in]{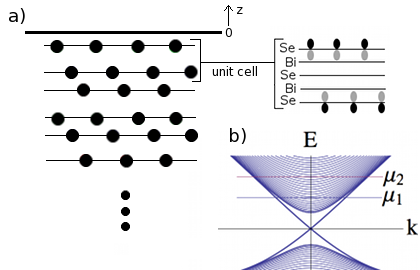}
\caption{a) Side view of a semi-infinite crystal of Bi$_{2}$Se$_{3}$. The two relevant $p_z$ orbitals are shown in the zoom-in view of the QL unit cell. 
b) Bulk and surface bands of the tight-binding model for Bi$_{2}$Se$_{3}$.  $\mu_1$ and $\mu_2$ denote two chemical potentials where the surface states have, respectively, not merged and merged into the bulk bands.}
\end{figure}

Consider a semi-infinite  Cu$_x$Bi$_2$Se$_3$ crystal occupying $z<0$,  which is naturally cleaved between QLs (see Fig.1). 
The realistic boundary condition corresponding to such a termination in the continuum $k\cdot p$ theory is\cite{fuberg} 
\beq
\sigma_z \psi(z=0) = \psi(z=0) \label{bc}.  
\eeq
This boundary condition reflects the vanishing of the electron wavefunction on the bottom layer ($\sigma_z=-1$) at $z=0$. 
Solving the differential equation 
\beq
E \psi &=& H(k_x, k_y, -i\partial_z)\psi \label{ss}
\eeq
subject to (\ref{bc}), we find two branches of mid-gap states 
\beq
\psi_\pm(k_x, k_y,z) &=& e^{z/l } (1, 0)_\sigma \otimes ( 1, \pm ie^{i \phi})_s,  \label{sf}
\eeq
where $ l = - v_z/m$ is the decay length, $\phi$ is the azimuthal angle of $(k_x, k_y)$, and the subscripts $\sigma$ and $s$ denote the orbital $\sigma_z$ and spin $s_z$ basis.  For $v_z m >0$, there are no decaying solutions; only when $v_z m <0$ in (\ref{sf}) do we obtain surface states decaying in the $-z$ direction.  The dispersion of these surface states is $E_\pm(k_x, k_y) = \pm v \sqrt{k_x^2 + k_y^2} \equiv \pm v k$, which agree well with the photoemission data from Cu$_x$Bi$_2$Se$_3$\cite{wray}. 
Thus, the existence of surface states on surfaces terminated between QLs  establishes  $v_z m<0$ in $H(\bk)$ for Cu$_x$Bi$_2$Se$_3$\cite{detail}. 


Having established that $v_z m<0$ and $v$ parameterizes the linear dispersion of the surface states, we now turn to the superconducting state of Cu$_x$Bi$_2$Se$_3$. Ref.\cite{fuberg} classified four different pairing symmetries   
compatible with short-range pairing interactions, and found that a spin-triplet, orbital-singlet, odd-parity pairing symmetry 
is favored when the inter-orbital attraction exceeds the intra-orbital one. 
The mean-field Hamiltonian of this superconducting state is 
\beq
H_{\rm MF}& =& \int d \bk [c_{\bk}^\dagger, \bar{c}_{-\bk}] {\cal H}(\bk)  
\left[
\begin{array}{c}
c_{\bk} \\
\bar{c}_{-\bk}^\dagger
\end{array}
\right], \nonumber \\
{\cal H}(\bk)& =& ( H(\bk) - \mu) \tau_z  + \Delta \sigma_y s_z \tau_x. \label{bdg}
\eeq
Here $c_{\bk}^\dagger = (c_{\bk,1\uparrow}^\dagger, c_{\bk,1\downarrow}^\dagger, c_{\bk, 2\uparrow}^\dagger, c_{\bk,2\downarrow}^\dagger)$ and $\bar{c}_{-\bk} \equiv c_{-\bk} \cdot is_y$ are four-component electron operators, with the subscript $1, 2$ labeling the two orbitals (Fig.1a). 
In the Bogoliubov-de Gennes Hamiltonian ${\cal H}(\bk)$, $\tau_{x}$ and $\tau_z$ are Pauli matrices in Nambu space,  
$\Delta$ is the pairing potential, and $\mu>|m|$ is the chemical potential in the conduction band. 

The above odd-parity superconducting Cu$_x$Bi$_2$Se$_3$ is fully-gapped in the bulk but has topologically protected 
surface Andreev bound states.  To determine the wavefunction and dispersion of these bound states, we begin by solving the BdG Hamiltonian 
${\cal H}(k_x, k_y, -i \partial_z)$ for the SABS at $k_x=k_y=0$. We find a Kramers pair of $\epsilon=0$ 
eigenstates\cite{detail}:
\beq
 \psi_{k=0, \alpha} (z)&=& e^{z \cdot \Delta  / |v_z|}   (\sin(k_{F} z  - \theta), \sin(k_{F} z))_\sigma   \nonumber \\
& \otimes & \left[ (1, -\alpha)_s,  i {\rm sgn}(v_z) (1, \alpha)_s  \right]_\tau, \; \alpha=\pm 1 \label{wf}
\eeq
Here $k_{F} \equiv  \sqrt{\mu^2 - m^2 }/v_z $ is Fermi momentum in the $z$ direction, and 
$\theta$  is  defined by $e^{i \theta} = (m + i \sqrt{\mu^2 - m^2})/\mu$. The subscript $\tau$ denotes a Nambu spinor. 
The Bogoliubov quasiparticle at $k=0$ is defined by 
$
\gamma_{\alpha} = \int dz \; \psi_{k=0, \alpha}(z) (c^\dagger(z), \bar{c}(z))_{k=0}^T. 
$
It is straight-forward to verify that $\gamma_{ \alpha}^\dagger = \gamma_{\alpha}$ up to an unimportant overall phase. 
This means that such quasiparticles are two-component massless Majorana fermions in $2+1$ dimensions.  

Having found the SABS wavefunction at $\epsilon=0$, $k=0$, we now show that the SABS dispersion crosses $\epsilon=0$ again at ${\it finite}$ $k$, which is one of the main results of this paper.  We establish this second crossing in two different ways: first, by a direct calculation, and second, by a topological argument.  It will become evident that the two approaches yield complementary information.

In the direct approach, we search for a second crossing by asking for which $k_0>0$ does ${\cal H}(0,k_0,-i\partial_z)\psi=0$ have a solution (it suffices to consider $k_x=0, k_y \equiv k_0>0$ only, due to rotational invariance). We find that $k_0$ is the nontrivial solution of the algebraic equation\cite{detail} 
\beq
|x|^2 + 2 {\rm sgn}(v_z) \frac{  E_F}{m} {\rm Re}(x) - 1=0, 
\eeq 
where $x$ is defined as 
\beq
x \equiv \frac{vk_0-i(\Delta+iE_F)}{\sqrt{(vk_0)^2+(\Delta+iE_F)^2}}, \; E_F \equiv \sqrt{\mu^2 - m^2}.
\eeq  
For Cu$_x$Bi$_2$Se$_3$ in the normal state with $\Delta = 0$ and $v_z m<0$, the above equation has a solution $k_0= \mu/v$, which  exactly correspond to the topological insulator surface states at Fermi energy obtained earlier in (\ref{sf}). 
With superconductivity, topological surface states in the normal state turn into SABS, with their location $k_0$ and wavefunction $\psi_{k_0, \alpha}$ perturbed by $\Delta$: 
$k_0 \simeq \frac{\mu}{v}(1-\frac{\Delta^2}{2m^2})$ 
and $\psi_{k_0, \alpha}$  acquires particle-hole mixing to first order in $\Delta$. 
Due to rotational invariance of the $k \cdot p$ Hamiltonian, the second crossing, hereafter denoted by $k_0$, exists along all directions in the $xy$ plane.  
This leads to a Fermi surface of SABS.  

In the topological approach, we first solve for the SABS dispersion at small $k$ and use topological arguments to infer its behavior at large $k$.  
Again, we set $k_x=0$ for convenience.  
Treating the $k_y$-dependent term in $H_{\rm BdG}$ as a perturbation, we find the dispersion is linear near $k=0$: 
$\epsilon_\alpha(k) = \alpha \tilde{v} k + o(k^3)$, forming a Majorana cone.  
The velocity $\tilde{v}$ is given by:
\beq
\tilde{v} &=& v \frac{\Delta^2 + {\rm sgn}(v_z)\Delta m}{\Delta^2 + {\rm sgn}(v_z) \Delta m + \mu^2}  \simeq  v\cdot {\rm sgn}(v_z)  \frac{  m \Delta}{ \mu^2}.  \label{velocity}
\eeq
In the second equality, we have used the fact $\Delta \ll |m| <\mu$ for weak-coupling superconductors. 

In (\ref{velocity}), it is important that the SABS velocity $\tilde{v}$ at $k=0$ 
has an {\it opposite} sign from the band velocity $v$ in the normal state of the doped topological insulator 
Cu$_x$Bi$_2$Se$_3$ ($v_z m <0$). As we now show, this fact has crucial implications for 
the SABS dispersion away from $k=0$: the two branches of SABS $\psi_{k, \pm}$ must 
cross each other at $\epsilon=0$ an odd number of times between $\bar{\Gamma}$ and the surface Brillouin zone edge $\bar{M}$. 
The existence of such additional crossings is dictated by a topological invariant we call ``mirror helicity'', 
which is a generalization of mirror Chern number\cite{teofukane} in topological insulators to topological superconductors. 
To define this invariant, note that the crystal structure of Cu$_x$Bi$_2$Se$_3$ has a mirror reflection 
symmetry $x \rightarrow -x$. As a result, the band structure (\ref{kp}) is invariant under mirror. 
However, the pairing potential in (\ref{bdg}) changes sign under mirror reflection. So the BdG Hamiltonian is 
invariant under a mirror reflection combined with a $Z_2$ gauge transformation $\Delta \rightarrow -\Delta$: 
\beq
{\cal H}(k_x, k_y, k_z) &=& \tilde{M} {\cal H}(-k_x, k_y, k_z) \tilde{M}^{-1},
\eeq
Here $ \tilde{M}= M \tau_z$, $M=-i s_x$ represents mirror reflection on electron spin. 
Because of this generalized mirror symmetry, bulk states are grouped into two classes with mirror eigenvalues $\pm i$ respectively. 
Each class can have a nonzero Chern number $n_{\pm i}$.  
Time reversal symmetry requires $n_{+i} = - n_{-i}$. 
The magnitude $|n_{+i}|=|n_{-i}|$ determines the number of helical Andreev modes with $k_x=0$ on the edge of $yz$ plane, while the sign defines a $Z_2$ mirror helicity:  
$
\eta \equiv {\rm sgn} (n_{+i}) = - {\rm sgn} (n_{-i}). 
$
The bulk topological invariant $\eta$ determines the helicity of such Andreev modes.    
For instance, $\eta < 0$ implies that the mode with mirror eigenvalue $-i$($+i$) moves clockwise(anti-clockwise) with respect to $+x$ axis at the edge of the $yz$ plane, and its energy-momentum dispersion curve must eventually merge into the $E > 0$ bulk quasiparticle continuum at a large positive(negative) momentum.  
Similar bulk-boundary correspondence applies to surface states in topological insulators\cite{teofukane, murakami}. 

As we show in Supplementary Material\cite{detail}, 
the topological superconducting phase of Cu$_x$Bi$_2$Se$_3$ and the undoped topological insulator Bi$_2$Se$_3$ 
have the same mirror helicity $\eta$, which is determined by the sign of the Dirac band velocity $v$ in the bulk. 
Given the relation between $\eta$ and helicity of surface excitations, this implies that the SABS in Cu$_x$Bi$_2$Se$_3$ must have the same helicity as surface states in Bi$_2$Se$_3$.  
On the other hand, the SABS velocity $\tilde{v}$ at $k=0$ has an opposite sign from the Dirac band $v$. To reconcile this fact with the helicity requirement,    
the two SABS branches $\psi_{k, \alpha}$---which are mirror eigenstates with eigenvalues $\tilde{M}=i \alpha$---must become twisted and switch places before merging into the bulk.  This necessarily results in an odd number of crossings between $\bar{\Gamma}$ and $\bar{M}$.   

The above topological argument reveals the robustness of gapless SABS at the second crossing in the $k\cdot p$ regime and beyond.   
In the $k\cdot p$ regime, the surface states at $\bk$ and $-\bk$ have {\it opposite} mirror eigenvalues (or spins) due to their helical nature, 
whereas the pairing symmetry $\Delta$ only pairs states with the {\it same} mirror eigenvalues. This symmetry incompatibility makes the surface states 
remain gapless in the topological superconducting phase \cite{mirror}.  
Moreover,  the topological argument demonstrates that the second crossing is topologically protected by the mirror helicity invariant in the bulk, as long as $\tilde v/v<0$ at $k=0$. As a result, the second crossing remains in a much larger energy range, 
even when higher order corrections to the $k \cdot p$ Hamiltonian become important, as shown below.  
In particular, we emphasize that the existence of the second crossing is {\it independent} of whether surface states are separated from the bulk at the Fermi energy. 

To gain more insight into these twisted SABS and to calculate their local density of states, we explicitly obtain its dispersion in the entire surface Brillouin zone.  For this purpose, we construct a two-orbital tight-binding model in the rhombohedral lattice shown in Fig.1 and calculate the SABS dispersion numerically.  
Details of our tight-binding model and its distinction from previous models\cite{ando, haolee} are described in the Supplementary Material\cite{detail}.  

Here we would like to note the following aspects of our model.  The normal state tight-binding model is constructed to reproduce both the $k\cdot p$ Hamiltonian (1) of Cu$_x$Bi$_2$Se$_3$ in the small $k$ limit and the boundary condition (\ref{bc})  in the continuum theory.  The bulk and surface bands of the normal state tight-binding model are displayed in Figure 1b; at chemical potential $\mu_1$, the Fermi momentum is relatively small and terms higher order than $\bk$ are negligible, whereas at $\mu_2$, these higher order terms cause deviation from the $k\cdot p$ Hamiltonian.  

\begin{figure}
\centering
\includegraphics[width=3.5in]{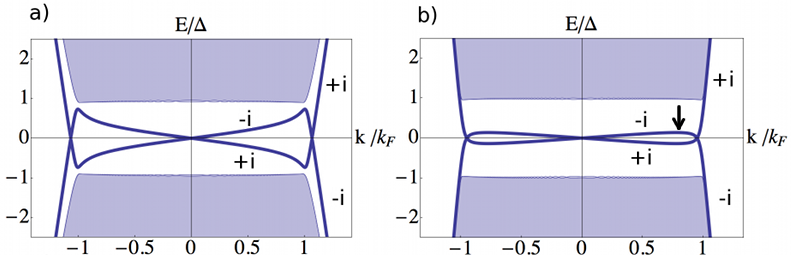}
\caption{SABS dispersion for the tight-binding model in which a) $m=-0.3<0$, $\mu_1=0.6$ and b) $m=-0.3<0$, $\mu_2=1$;   
The mirror eigenvalues are displayed near each branch of SABS. The SABS twist with a second crossing near Fermi momentum, as also observed in Ref.\cite{haolee}.  The arrow denotes where the dispersion has zero slope, resulting in a Van Hove singularity in the density of states.}
\end{figure}

Upon adding odd-parity superconductivity pairing to the model, we obtain the SABS dispersion (Fig. 2).  A branch of linearly dispersing Majorana fermions is found at $k=0$, which signifies a three-dimensional topological superconductor. In addition, the bands of Andreev bound states in the surface Brillouin zone are twisted:  they connect the Majorana fermion at $k=0$ with 
the second crossing near Fermi momentum.  Such behavior was independently found by Hao and Lee\cite{haolee, detail}, 
and its topological origin is revealed by our analytical calculations and arguments.   

For a given branch ($\tilde{M}= \pm i$) of SABS, its particle-hole character evolves as a function of momentum from having an equal amount of particle and hole (charge neutral) at $k=0$ to being 
exclusively hole or particle (charged) at large $k$. 
At chemical potential $\mu_1$, the SABS near the second crossing can be identified with nearly unpaired surface states in the normal state, which show up twice---as particle and hole---in the BdG spectrum.  
However, even when these surface states have merged into the bulk, the SABS still has the second crossing, as required by the mirror helicity.  This is shown in Fig. 2b, at chemical potential $\mu_2$. The resulting gapless SABS near the second crossing has substantially more particle-hole mixing than the first case 
and is unrelated to surface states in the normal state.  Such SABS defy a quasi-classical description 
and represent a new type of Andreev bound states which arises from the interplay between nontrivial band structure and unconventional superconductivity. 

Finally, we relate our findings of SABS in Cu$_x$Bi$_2$Se$_3$ to the recent point-contact spectroscopy experiment\cite{ando},  in which a zero-bias differential conductance peak along with a dip near the superconducting gap edge was observed below $1.2$K  and attributed to SABS. To compare with this experiment, we calculate the local tunneling density of states (LDOS) as a function of energy for $m/\mu_2=0.3$---roughly the value found in ARPES\cite{wray}.    
The resulting LDOS at zero and finite temperatures are shown in Fig. 3.  The finite temperature LDOS from $T=0.05\Delta$ to $T=0.2\Delta$ agrees with the experimentally observed differential conductance peaks as well as the dips with the slight asymmetry between positive and negative voltages.  
Both features along with the absence of coherence peaks contrast sharply with the tunneling spectrum of an s-wave superconductor.  

In addition to comparison with the experiment, we make the following predictions stemming from the zero temperature LDOS in Figure 3a.  Here the two peaks arise from Van Hove singularities at the particular energy near $E=0$ where the SABS bands have zero slope, indicated by the arrow in Fig. 2b.  Furthermore, the significant asymmetry in the height of these two peaks reflects the fact that the SABS at the turning point is primarily of hole type, as noted earlier. 
The energy of these two peaks and the magnitude of their asymmetry depends somewhat on details of band structure.  
However, the existence of two peaks only depends on there being a turning point in the SABS dispersion, which is guaranteed by the existence of a second crossing in a wide regime of chemical potentials. 
Hence, we predict that for relatively clean surfaces the zero-bias conductance peak in the tunneling spectra will split into two asymmetric peaks at even lower temperatures. 
Such peaks will be an unambiguous signature of Majorana fermions smoothly turning into normal surface electrons.   
Furthermore, the SABS dispersion we predict in Fig.2  can be directly tested in future ARPES experiments.

\begin{figure}
\centering
\includegraphics[width=3.5in]{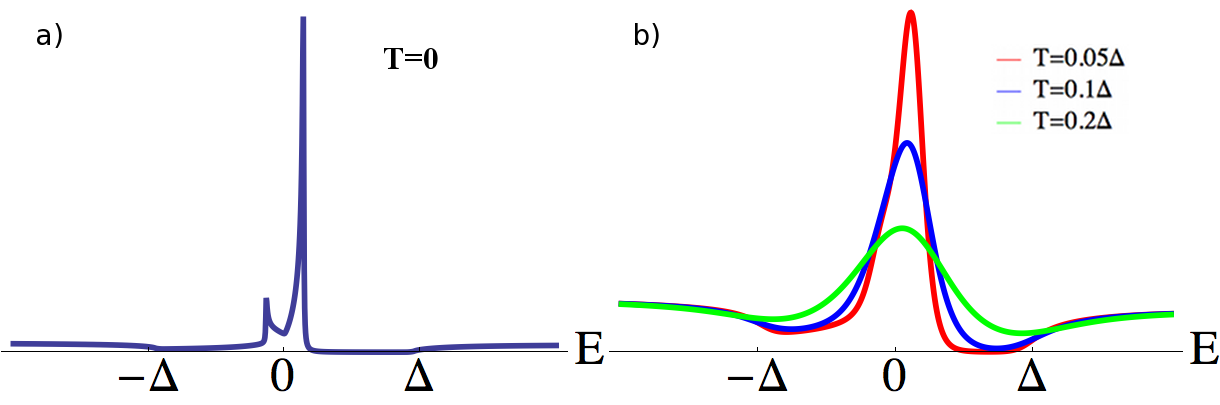}
\caption{Tunneling local density of states (arbitrary units) at a) $T=0$ and b) finite temperature.  In both cases, the chemical potential is $\mu_2=1$.}
\end{figure}

While the main focus of this Letter is Cu$_x$Bi$_2$Se$_3$, we end by discussing the implications of our findings for superconducting doped semiconductors with similar band structures. Candidates include Bi$_2$Te$_3$\cite{bite} under pressure, TlBiTe$_2$\cite{thallium}, PbTe\cite{pbte}, SnTe\cite{snte}, and GeTe\cite{gete}. Provided that the material is inversion symmetric and its Fermi surface is centered at time-reversal-invariant momenta, the Dirac-type relativistic $k\cdot p$ Hamiltonian (\ref{kp}) describes their band structures\cite{fukane}. Moreover, if the pairing symmetry is odd under spatial inversion and fully gapped, the system is (almost) guaranteed to be a topological superconductor according to our criterion\cite{fuberg, qifu}.  Our work is also relevant to noncentrosymmetric superconductors such as YPtBi\cite{yptbi}, if their pairing symmetries have dominant odd-parity components.   

As a final point which captures the essence of this work, we compare and contrast SABS in doped superconducting topological insulators with normal insulators, which differ by a band inversion ($v_z m <0$ versus $v_z m>0$).  In both, the Majorana fermion SABS exist at $k=0$ as shown in (\ref{wf}, \ref{velocity}). 
However, the SABS in doped normal insulators do not necessarily have the second crossing near Fermi momentum\cite{detail}. This can be understood from our mirror helicity argument, with the difference being that $\tilde{v}/v >0$ for $v_zm>0$ (see Eq.(\ref{velocity})).  
In this sense, the new type of surface Andreev bound state and its phenomenological consequences are the unique 
offspring of both nontrivial band structure and odd-parity topological superconductivity.   

{\it Note:}  Two recent studies\cite{haolee, ando}  calculated the surface spectral function numerically in Cu$_x$Bi$_2$Se$_3$ tight-binding models.   
The second crossing of SABS was independently found in Ref.\cite{haolee}. 
We also learned of another point-contact measurement on Cu$_x$Bi$_2$Se$_3$\cite{point}. 

{\it Acknowledgement}: We thank Yoichi Ando, Erez Berg, Chia-Ling Chien, Patrick Lee and Yang Qi for helpful discussions, as well as Anton Akhmerov and David Vanderbilt for helpful comments on the manuscript. This work is supported by DOE under cooperative research agreement Contract Number DE-FG02-05ER41360 and NSF Graduate Research Fellowship under Grant No. 0645960 (TH), the Harvard Society of Fellows and MIT start-up funds (LF).
LF would like to thank Institute of Physics in China, and Institute of Advanced Study at Tsinghua University for generous hosting.

\section{Supplementary Material}

\subsection{I. SABS Wavefunction and Second Crossing}

First, we derive in detail the wavefunction (6) in the main text from the BdG Hamiltonian ${\cal H}(\bk)$.  A Kramers pair of zero-energy eigenstates $\psi_{k=0, \alpha=\pm}(z)$ with mirror eigenvalues $\tilde{M}= i \cdot \alpha$ is expected from the topology and symmetry of ${\cal H}(\bk_\parallel=0)$. 
For a given mirror eigenstate, $s_x$ is locked to $\tau_z$ by the identity $s_x \tau_z = - \alpha$, so that  
$\psi_{k=0, \alpha}(z)$ satisfies a reduced $4$-component equation:
\beq
 [ (m \sigma_x -i v_z \sigma_y \partial_z   - \mu) \tau_z + \Delta \sigma_y \tau_x ] \psi_{k=0, \alpha}(z) =0 .  \label{sfeq1}
\eeq 
This can be further simplified by multiplying both sides by $\tau_z$: 
\beq
 [ m \sigma_x -i v_z \sigma_y \partial_z   - \mu +   i  \Delta \sigma_y \tau_y  ] \psi_{k=0, \alpha}(z) =0 . \nonumber \label{sfeq2}
\eeq
It is evident that $\psi_{k=0,\alpha}$ is an eigenstate of $\tau_y$. 
The corresponding eigenvalue is given by ${\rm sgn }(v_z) $ in order to have a decaying solution. 
Eq.(\ref{sfeq1}) then reduces to a two-component equation in orbital space, which has two independent solutions:
\beq
\xi_\pm(z) = (1, e^{\pm i \theta})_\sigma \cdot e^{(\pm i k_{F} + \Delta/ |v_z|)  z}. \label{xi}
\eeq
$\theta$  is defined by $e^{i \theta} = (m + i \sqrt{\mu^2 - m^2})/\mu$. 
Choosing a suitable linear combination of $\xi_+$ and $\xi_-$ to satisfy the boundary condition (2) in the main text, we obtain 
the wavefunction of SABS, which is reproduced here for the reader's convenience:
\beq
\psi_{k=0, \alpha} (z)&=& e^{z \cdot \Delta  / |v_z|}   (\sin(k_{F} z  - \theta), \sin(k_{F} z))_\sigma   \nonumber \\
& \otimes & \left[ (1, -\alpha)_s,  i {\rm sgn}(v_z) (1, \alpha)_s  \right]_\tau, \; \alpha=\pm 1 .  \nonumber 
\eeq

Next we solve for the location of the SABS second crossing.  For convenience, we look for a zero-energy solution $\psi(z)$ at $k_x =0$, $k_y\equiv k_0$ with mirror eigenvalue $+i$ (i.e., $s_x  \tau_z=-1$). $\psi$ satisfies 
\beq
\left[ (m\sigma_x - i v_z \sigma_y \partial_z   - \mu ) \tau_z  + v k_0 \sigma_z  + \Delta \sigma_y  \tau_x \right] \psi(z) = 0 \label{maineq}
\eeq 
Recall that $v_z m <0$ for a doped topological insulator. Without loss of generality, here we choose $m<0, v_z>0$.  
By multiplying Eq.(\ref{maineq}) by $i \sigma_y \tau_z$, the zero-energy solution satisfies
\beq
\left[ m\sigma_z  + v_z  \partial_z   - i \mu \sigma_y  - v k_0 \sigma_x \tau_z - \Delta  \tau_y \right] \psi(z) = 0. \label{eq}
\eeq
We write the wavefunction $\psi(z)$ as 
\beq
\psi(z) = e^{  i \lambda \sigma_x /2} \phi(z), \; \lambda \in \rm C \label{rotate}
\eeq
where $\cos \lambda=\mu/E_F$ and $\sin \lambda = - i m/E_F$, $E_F \equiv \sqrt{\mu^2 - m^2}$.  
Eq.(\ref{maineq}) then becomes
\beq
\left[  v_z  \partial_z   - i E_F \sigma_y  - v k_0 \sigma_x \tau_z - \Delta  \tau_y \right] \phi(z) = 0 \label{eq}
\eeq
Note that Eq.(\ref{eq}) now commutes with $\sigma_y \tau_y$, which becomes a constant labeled by $\tau$. The reduced equation 
for $\phi_\tau(z)$ is  
\beq
\left[  v_z  \partial_z   -(\tau \Delta + i E_F )\sigma_y  - v k_0 \sigma_x  \right] \phi(z) = 0 \label{reduced}.
\eeq
The solution takes the form $\phi(z)=e^{Kz}\xi$.  
First consider $\tau=1$. From Eq. (\ref{reduced}), we have 
\beq
K_{\pm}=\frac{\pm \sqrt{(vk_0)^2 + ( \Delta + i E_F)^2 }}{v_z} \equiv \pm E_z /v_z.
\eeq
Corresponding eigenvectors are given by
\beq
\xi_{\pm}=( x_\pm, 1), \; x_\pm \equiv (vk_0 - i ( \Delta + i E_F)) / (v_z K_\pm)
\eeq
To get a decaying solution,  we must have ${\rm Re}(K)>0$.      
Hence, we must choose $K_+$ and thus $\xi_+$.
We now rewrite the complete wavefunction with both orbital and Nambu components (spin is locked by $s_x \tau_z=-1$ and not shown explicitly):
\beq
\xi_{+} &=& \frac{x_+-i}{2}(1,i,i,-1) + \frac{x_++i}{2}(1,-i,-i,-1) \nonumber \\
&=& (x_+, 1, 1, -x_+)
\eeq
Note that the equation for $\tau=-1$ is equivalent to the complex conjugate of that for $\tau=1$. Therefore  
if we choose $K_+ \equiv K$ and $x_+ \equiv x$ for $\tau=1$, we must choose $K^*$ and $x^*$ for $\tau=-1$.  The corresponding wavefunction is 
\beq
\xi^*=(x^*, 1, 1, -x^*)
\eeq
It follows from Eqn. (\ref{rotate}) that 
\beq
\psi(\tau=1) &=& e^{i \lambda \sigma_x/2}(x, 1, 1, -x) \nonumber \\
\psi(\tau=-1) &=& e^{i \lambda \sigma_x/2}(x^*, 1, -1, x^*)
\eeq
Up to normalization, the most general form of $\psi(z)$ satisfying the boundary condition (\ref{bc}) is
\beq
\psi(0)=(1,0,A,0),
\eeq
where $A$ is some constant.
Hence, for a nontrivial solution to exist, the determinant of the $2\times 2$ matrix made from the second and fourth component of $\psi(\tau=1)$ and $\psi(\tau=-1)$ 
must be zero. This condition is simplified to an algebraic equation 
\beq
0 &=& 2{\rm Re}(x) +\frac{m}{E_F} (-1 + |x|^2) \label{result}
\eeq
which is the result cited in the main text.
Our previous solution at $k=0$ (\ref{wf}) corresponds to $x=\pm i$, which satisfies the above condition.  
Another simple limit is the normal state with $\Delta=0$. In this case, the second crossing is simply located at the momentum 
where the topological insuator surface states cross the chemical potential, namely, $k_0=\mu/v$.  We can check that for this case, $x=(\mu+E_F)/(-m)$ indeed satisfies Eq. (\ref{result}).
Now we solve for $k_0$ to first order in $\Delta$.  
Temporarily absorbing $v$ into $k_0$ and expanding $x$ to second order in $\Delta$, we obtain
\beq
{\rm Re}(x)&=&\frac{E_F+k_0}{\sqrt{-E_F^2+k_0^2}} \nonumber \\
&&+\frac{(-2E_F k_0 -k_0^2)\Delta^2}{2(E_F-k_0)^2(E_F+k_0)\sqrt{-E_F^2+k_0^2}} \nonumber \\
{\rm Im}(x)&=&\frac{k_0\Delta}{(E_F-k_0)\sqrt{\-E_F^2+k_0^2}} \nonumber \\
\frac{m}{E_F}(|x|^2-1)&=&\frac{-2m}{E_F-k_0}+\frac{2m k_0 \Delta^2}{(E_F-k_0)^3(E_F+k_0)} \nonumber
\eeq
From Eq.(\ref{result}), we then extract the leading order to correction to $k_0$:
\beq
k_0=\frac{\mu}{v}(1-\frac{\Delta^2}{2m^2}).
\eeq
The corresponding $x$ at $k_0$ is given 
\beq
{\rm Re}(x)&=&\frac{m}{-\mu+E_F}-\frac{\Delta^2 \mu (\mu + E_F)}{2m^3 (-\mu+E_F)} \\
{\rm Im}(x)&=&-\frac{\Delta \mu}{m(-\mu+E_F)}
\eeq
We conclude this section by calculating the ratio of the particle ($\tau=1$) and hole ($\tau=-1$) components of the $s_x\tau_z=- 1$ wavefunction $\psi(z)$ at the second crossing and at $z=0$.  This wavefunction is some linear combination $c_1 \psi(\tau=1)+c_2(\tau=-1)$ with vanishing second and fourth components (to satisfy the boundary condition).   
Hence,  we find 
\beq
\frac{c_2}{c_1}=\frac{-(\cos(\lambda/2)+i x\sin{\lambda/2})}{\cos{\lambda/2}+i x^* \sin{\lambda/2}}
\eeq
The hole/particle ratio is
\beq
r\equiv\frac{\cos{\lambda/2}(c_1-c_2)-i\sin{\lambda/2}(c_1 x-c_2 x^*)}{\cos(\lambda/2)(c_1 x+c_2 x^*)+i\sin(\lambda/2) (c_1 + c_2)}
\eeq
Using the fact that the second and fourth components vanish, which is equivalent to 
\beq
\cos{\lambda/2}(c_1+c_2)+i\sin{\lambda/2}(c_1 x+c_2 x^*)=0 \\
-\cos{\lambda/2} (c_1 x-c_2 x^*)+i\sin{\lambda/2}(c_1-c_2)=0,
\eeq
we get
\beq
r=\frac{c_1-c_2}{(c_1+c_2)(i \cot(\lambda/2))}=\frac{1+i(\tan{\lambda/2} ){\rm Re}(x)}{i {\rm Im}(x)}
\eeq 
Recalling that $\cos \lambda=\mu/E_F$ and $\sin \lambda = - i m/E_F$, we have 
\beq
\tan(\lambda/2)=\frac{e^{i\lambda}-1}{i(e^{i\lambda}+1)}=\frac{\mu+m-E_F}{\mu+m+E_F}
\eeq
The hole/particle ratio at the second crossing is thus
\beq
r=\frac{\Delta(\mu+E_F)(\mu+m-E_F)}{2im^2(\mu+m+E_F)}, 
\eeq
which is first order in $\Delta$. 

\subsection{II. Mirror Helicity}
Here we show that the topological insulator and topological superconductor phases have the same mirror helicity.
We  deduce this fact from the phase transition between topological insulators and topological superconductors. 

The BdG Hamiltonian (5) in the main text exhibits three topologically distinct gapped phases as a function of the band gap, pairing potential and doping. 
At zero doping ($\mu=0$) and in the absence of superconductivity ($\Delta=0$), 
the system is either an normal band insulator or a topological insulator, depending on the sign of $m$.  
At finite electron doping, the chemical potential lies inside the conduction band: $\mu>0$. 
When the odd-parity pairing $\Delta$ occurs in such a doped normal insulator or topological insulator, the system becomes a fully gapped topological superconductor. 
For the sake of our argument,we note that the topological superconductor phase is adiabatically connected to the $\mu=0$ and $\Delta > |m|$ limit. 
Fig.4 shows the three phases in the $\mu=0$ phase diagram as a function of $m$ and $\Delta$.  
The phase transition between topological superconductors and normal/topological insulators occurs at $\Delta=\pm m$. 

\begin{figure}
\centering
\includegraphics[width=3in]{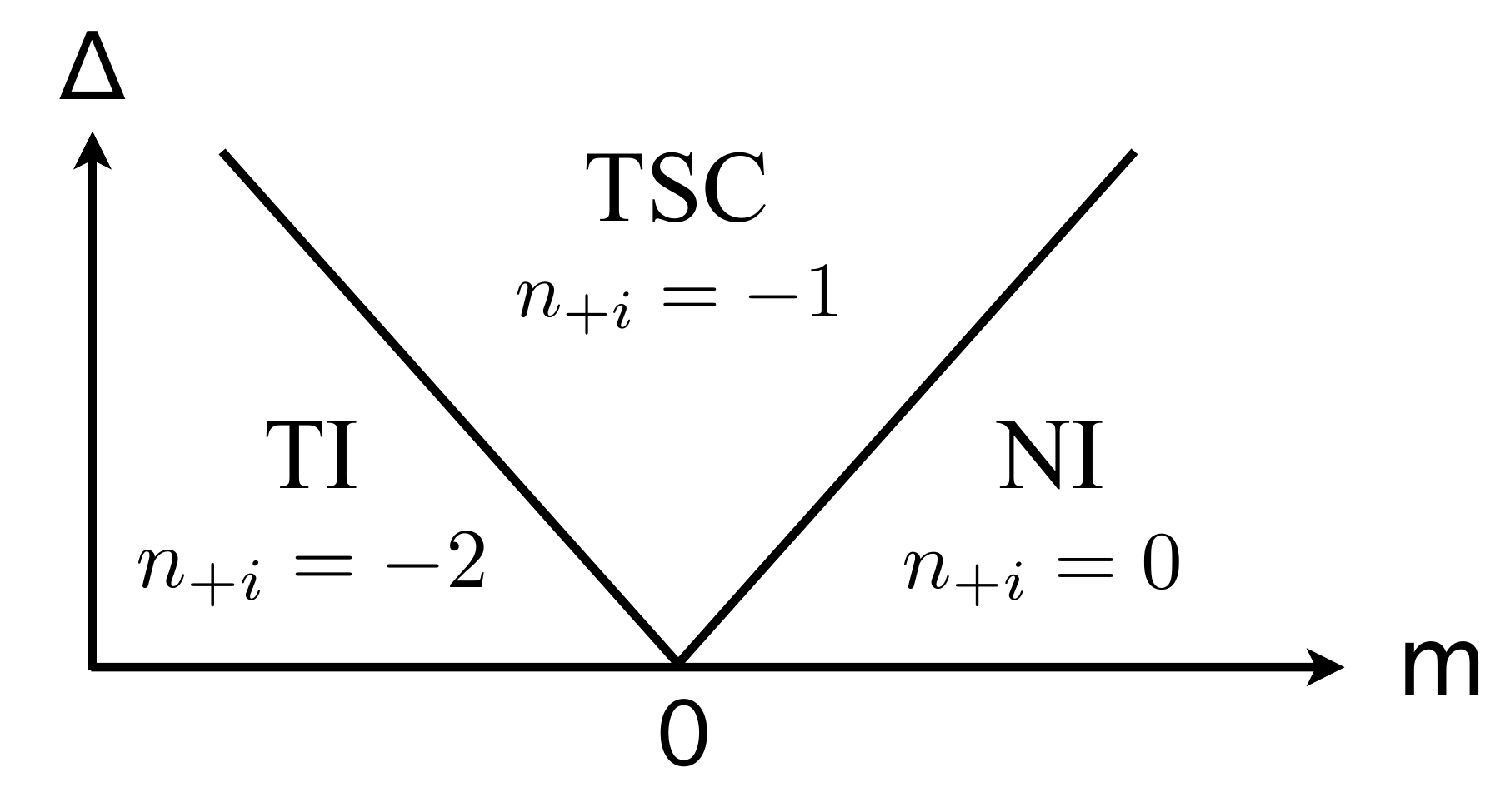}
\caption{Phase diagram of fully-gapped odd-parity superconductivity in doped semiconductors as a function of band gap $m$ and pairing potential $\Delta$, showing three gapped phases: band insulator, topological insulator and topological superconductor. 
They are topologically distinguished by the mirror Chern numbers $n_{+i}$. }
\end{figure}

Recall from the main text that due to mirror symmetry, each phase has a mirror Chern number $n_{+ i} = n_{-i}$ displayed in Fig. 4.  
Using $n_{+i}=0$ for the normal insulator as a reference, we can obtain the mirror 
Chern number for the topological insulator and topological superconductor  
by calculating the change of $n_{+i}$ across the phase transition to the normal insulator. Due to the double counting of particles and holes, the mirror Chern number of a band insulator defined in Nambu space is always an even integer twice the value of that defined previously for insulators in Ref.[25]. 
As a result,  a direct transition from topological insulator to band insulator at $\Delta=0$  changes $n_{+i}$ by two. For $\Delta \neq 0$, this transition is split into two transitions with an intermediate topological superconductor phase, so that each transition changes $n_{+i}$ by one.  
Therefore we have
 \beq
 n_{+i}({\rm TI}) = 2 n_{+i}({\rm TSC}).  \label{mc}
 \eeq 
Recall that mirror helicity is defined as $\eta\equiv {\rm sgn}(n_{+i})$.  Hence, the topological insulator and topological superconductor phase have the same mirror helicity.

\subsection{III. Tight-binding Model} 
Here we present the details of our tight-binding model. This model is defined on the rhombohedral lattice with a bilayer unit cell shown in Fig.1. The Hamiltonian   
$
H=H_{0}+H_{12}+ H_{\rm soc} + H'_{12} 
$
consists of the following four terms. 
\beq
H_{0}= \sum_{<ij>}  t_{0}c_{i\alpha}^{\dagger}c_{j\alpha} \nonumber 
\eeq
describes  nearest neighbor hopping within the same layer. 
\beq
H_{12}=\sum_{<i\in1,j\in2>}t_{1}c_{i\alpha}^{\dagger}c_{j\alpha} + \sum_{<i\in1,j'\in2>} 
t_{2} c_{i\alpha}^{\dagger}c_{j'\alpha} 
\eeq
 describes hopping between two adjacent layers within a QL ($t_1$) and on two neighboring QLs ($t_2$).   
 $t_0$, $t_1$ and $t_2$ are spin-independent. 
In addition, the two orbitals in the upper and lower part of the unit cell (Fig.1a) experience local electric fields along the $\pm z$ direction, which give rise to  
the following Rashba-type spin-orbital associated with intra-layer hopping: 
\beq
H_{\rm soc}&=&(\sum_{<ij>\in 1} - \sum_{<ij>\in 2}) \frac{i \lambda}{2}c_{i\alpha}^{\dagger}{\vec s}_{\alpha\beta}c_{j\beta}\cdot(\hat{z}\times \mathbf{a}_{ij}), \nonumber
\eeq
where $\mathbf{a}_{ij}=\frac{1}{2}\epsilon_{ijk}(\mathbf{R}_j-\mathbf{R}_k)$ denote the vectors joining nearest neighbors within a layer, and $\mathbf{R}_{1,2,3}$ are the Bravais lattice vectors. The last term $H_{12}'$ (which plays a minor role) describes inter-layer second nearest neighbor ($t_{3})$ hopping within a QL:
$
H'_{12}= \sum_{<<i\in1,j\in2>>}t_{3}c_{i\alpha}^{\dagger}c_{j\alpha} + h.c. \nonumber
$

We emphasize that our tight-binding Hamiltonian $H$, by construction, satisfies the symmetries  of the Bi$_2$Se$_3$ crystal structure. 
Its point group $D_{3d}$ has three independent symmetry operations: 
 inversion $P$, three-fold rotation $C_3$ around the $z$ axis,  and
reflection $M$ about the $x$ axis. These operations act on the orbital and spin degrees of freedom as follows:
$P$ interchanges the two orbitals (see Fig.1a), $C_3$ rotates the electron spin $s_x$ and $s_y$, and
$M$ flips $s_z$ and $s_y$, but not $s_x$ (Recall that spin is a pseudovector). Therefore,  these operations are represented by 
 $P=\sigma_x, C_3= \exp(-i \frac{1}{2} \frac{2\pi }{3}s_z ), M= - i s_x$. 
  
The above tight-binding model captures the essential features of the band structure Bi$_2$Se$_3$ near the $\Gamma$ point. 
(We caution the reader that our tight-binding model does {\it not} aim to describe the band structure of Cu$_x$Bi$_2$Se$_3$ in the entire Brillouin zone. 
Such a task requires a realistic band structure modeling beyond the scope of this work.) 
First, the Bloch Hamiltonian $H(\bk)$ reduces to the $k\cdot p$ Hamiltonian (1)  in the main text 
as follows: 
$m=3(t_1+t_2+t_3)$, $v_z=3t_2 c$, and $v=\frac{9}{2}\lambda a^2$, where $a=|\mathbf{a}_{ij}|$ and $c=|\frac{1}{3}(\mathbf{R}_1+\mathbf{R}_2+\mathbf{R}_3)|$. 
Second, our model is able to reproduce the Dirac surface states (Fig.1b). To understand this, we note that 
at $k_x=k_y=0$, the spin-orbit term $H_{\rm soc}$ vanishes. The resulting one-dimensional system corresponding to 
$H(k_x=k_y=0)$ is equivalent to the well-known Su-Heeger-Schrieffer model for polyacetylene, which has a similar two-site unit cell. 
In both systems, the hopping between neighboring sites within a unit cell is different from that between two unit cells. As a result, when such a one-dimensional 
system is terminated on a ``strong bond'', zero-dimensional end states appear within the band gap and are spin degenerate. 
In contrast, when the system is terminated on a ``weak bond'', end states are absent. 
In the context of Bi$_2$Se$_3$, strong bond correspond to termination between two QLs, and weak bond correspond to termination within a QL. 
In the former case, the end states at $k_x=k_y=0$ disperse and become spin-split as a function of $k_x$ and $k_y$, due to the $k$-linear spin-orbit term $H_{\rm soc}$.    
As a result, they constitute the two-dimensional Dirac surface band of  Bi$_2$Se$_3$.   
In the latter case, the end states are absent at $k_x=k_y=0$. Instead, the surface state Dirac points of Bi$_2$Se$_3$ 
are located at three $\bar{M}$ points\cite{teofukane, termination} (which cannot and should not be accessed by 
$k\cdot p$ Hamiltonian near $\Gamma$).    
It will be interesting to experimentally verify such a drastic dependence of surface band structure on surface terminations.   

To capture the effect of two different surface terminations within a continuum  theory, 
we choose the boundary condition correspondingly.  
The boundary condition for termination between two QLs (strong bond) is 
\beq
\sigma_z \psi(z=0) = \psi(z=0) \label{bc}.  
\eeq
This reflects the vanishing of the $\sigma_z = -1$ component of the wavefunction  at $z = 0$ (the outmost site corresponds to $\sigma_z=1$). 
Instead, the boundary condition for termination with a QL  (weak bond) is 
\beq
\sigma_z \psi(z=0) = - \psi(z=0) \label{bc}.  
\eeq
As we have shown in the main text, for $v_z m<0$ Dirac surface states exist in the continuum theory for the first termination, but not for the second.  
This correctly reproduces the experimental phenomenology. 

To include superconductivity, we add the following odd-parity pairing term in the Hamiltonian:
\beq
H_{\rm MF} = H + \sum_{<i\in 1,j\in 2>} \frac{\Delta}{6} (c^\dagger_{i\uparrow}c^\dagger_{j\downarrow} + c^\dagger_{i\downarrow}c^\dagger_{j\uparrow} ) + h.c . \nonumber
\eeq 

The parameters we used are $\Delta=0.03, t_0=-0.1,t_1=-1,t_2=0.5,t_3=0.6, a=1, c=1, \lambda=0.5$, and $\mu=0.6$ (above the normal state surface Dirac point) for Figure 2a and $\mu=1$ (above the Dirac point) for Figure 2b.  The slab size was 320 unit cells.  We note that $v_z\propto t_2 >0$ actually corresponds to $v_z<0$ in the $k\cdot p$ Hamiltonian above because our simulated crystal is oriented in the opposite z direction relative to the $k\cdot p$ definition.

For completeness, we calculate the SABS dispersion for a doped band insulator ($m v_z>0$), in which the second crossing does not exist because $\tilde{v}/v>0$ (Fig. 5a).  The dispersion for the critical case ($m=0$) is displayed in Fig. 5b.

\begin{figure}
\centering
\includegraphics[width=3in]{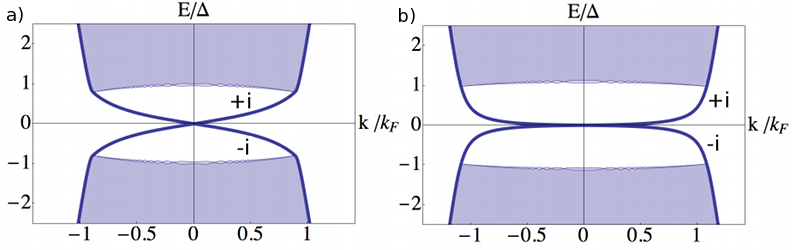}
\caption{SABS dispersion for the tight-binding model in which a) $m=0.3>0$, $\mu_1=0.6$ corresponds to a doped BI; b) $m=0$, $\mu_1=0.6$ corresponds to a doped zero-gap semiconductor.}
\end{figure}

\subsection{IV. Relation to Previous Works} 

In a recent work, Hao and Lee\cite{haolee} calculated the surface spectral function in tight-binding models for Cu$_x$Bi$_2$Se$_3$ with the four possible 
pairing symmetries\cite{fuberg}, including the fully-gapped odd-parity pairing studied in this work. They used two tight-binding models which are 
lattice regularizations of two $k\cdot p$ Hamiltonians (I and II). However, both these Hamiltonians violate the mirror symmetry. Model II is quoted from  
the incorrect $k\cdot p$ Hamiltonian of Ref.\cite{zhang}:  their terms $k_z \sigma_x s_z$ as well as  
$\sigma_x (k_x  s_x + k_y s_y)$, in the basis they {\it specify},  violates the mirror symmetry $M$. 
 A corrected version\cite{liu} is identical to our $k\cdot p$ Hamiltonian (1) 
after interchanging $\sigma_x$ and $\sigma_z$ (corresponding to a change of basis for the orbitals).  
Model I is claimed to be quoted from Ref.\cite{fuberg} (the one we use here). However, the term 
$\sigma_z (k_x s_y - k_y s_x)$ is mistakenly replaced by $\sigma_z (k_x s_x + k_y s_y)$.  

Nonetheless, if one forgoes the definition of $s_{x,y}$ as operators corresponding to spin along the x and y directions in real space, then their Model I corresponds to our $k\cdot p$ Hamiltonian after a unitary spin rotation $\exp(-i \frac{\pi}{4} s_z )$ (without affecting the odd-parity pairing term $\Delta \sigma_y s_z \tau_x$).  Hence, they also found that Majorana fermion Andreev bound states at $k=0$ connect to the Dirac surface states near Fermi momentum.  They attributed the second crossing to the fact that Dirac surface states remain gapless in the odd-parity superconducting state, and concluded that 
it disappears if the surface states merge into the bulk. In contrast, our work revealed the topological origin of the twisted surface Andreev bound states: as long as $\tilde{v}/v<0$, they are protected by mirror symmetry and exist independent of whether Dirac surface states appear at Fermi energy.

\subsection{V. Finite Temperature Differential Conductance}

Finally, we elaborate on how we attained the differential conductance plots in the main text.  
Consider two systems separated by an insulating barrier.  
Then the tunneling current is proportional to the transition rate given by Fermi's golden rule:
\beq
I\propto \int d \epsilon A^+_1( \epsilon + eV ) A^-_2(\epsilon)  - A^-_1(\epsilon+eV) A^+_2(\epsilon) 
\eeq
where $A^+(\epsilon)$ is the probability of adding a particle and changing the system's energy by $\epsilon$ (positive or negative), 
and $A^-(\epsilon)$ is the probability of removing a particle and changing the system's energy by $-\epsilon$.  $1$ and $2$ denote the two sides of the barrier. 

For free electron systems, $A^\pm$ is given by the density of states weighted by the Fermi-Dirac distribution
\beq
A^\pm(\epsilon) &=& \int d k A^{\pm}(\epsilon, k) \nonumber \\
A^{-}(\epsilon, k) &=& n_F(\epsilon) \delta(\epsilon - \xi_k) \nonumber \\
A^+(\epsilon, k) &=& (1-n_F(\epsilon) ) \delta(\epsilon - \xi_k) \label{free}
\eeq
where $n_F(\epsilon)$ is the Fermi-Dirac distribution function $1/(e^{\epsilon/T}+ 1 )$. 
For convenience, hereafter both $\epsilon$ and $\xi_k$ are measured with respect to chemical potential. 

For a BCS superconductor, $A^\pm$ is modified:
\beq
A^{-}(\omega, k) &=&  |u_k|^2 n_F(\omega) \delta(\omega - |\xi_k|), \; \omega>0 \nonumber \\
& &   |v_k|^2 (1-n_F(|\omega|)) \delta(|\omega |- |\xi_k|) , \; \omega< 0 \nonumber \\
A^+(\omega, k) &=&  |u_k|^2 (1-n_F(\omega)) \delta(\omega - |\xi_k|) , \; \omega>0 \nonumber \\
& &  |v_k|^2 n_F(|\omega|) \delta(|\omega| - |\xi_k|) , \; \omega<0  \label{bcs}
\eeq 
where $|\xi_k|>0$ is the energy cost of creating a quasi-particle excitation. $u$ and $v$ are the particle and hole components of the positive-energy eigenstates 
of BdG Hamiltonian, respectively.  To derive (\ref{bcs}), one must keep in mind that adding(removing) a quasi-particle always 
increases(decreases) the energy of the system. Because the hole component of a $E>0$ eigenstate is related to the particle component of its partner at $-E$ by 
the inherent particle-hole symmetry in BdG formalism, (\ref{bcs}) can be simplified to 
\beq
A^{-}(\omega, k) &=&  |u_k|^2 n_F(\omega) \delta(\omega - \xi_k),  \nonumber \\
A^+(\omega, k) &=&   |u_k|^2 (1- n_F(\omega)) \delta(\omega - \xi_k) ,  \label{bcs}
\eeq
where we have used $1- n_F(- \omega) = n_F(\omega)$. Here $\omega$ and $\xi_k$ can be both positive and negative. 
Written in this form, $A^\pm$ for a superconductor is similar to a normal metal, except it has  prefactor $u_k$. 
When superconductivity vanishes, $u_k=1, v_k=0$ for $k>k_F, \omega>0$ and $k<k_F, \omega<0$, whereas 
$u_k=0, v_k=1$ for $k<k_F, \omega>0$ and $k>k_F, \omega<0$. In this limit,  (\ref{bcs}) reduces to the free fermion case (\ref{free}).  

In our simulation, $|u_k|^2$ and $|v_k|^2$ were obtained from the $\tau=1$ and $\tau=-1$ components of the surface Green's function, summed over spin and for the $\sigma_z=+1$ orbital at $z=0$ only, in accordance with our boundary condition.  The surface Green's function was computed using a recursive algorithm \cite{greens}, allowing us to use a very large slab size ($2^{24}$ layers). 

Substituting $A^\pm$ into the expression for tunneling current and assuming that the density of states of the normal metal is constant, 
we obtain
\beq
I  &\propto& \int_{-\infty}^{\infty} d \epsilon \rho_N  \rho_S(\omega) [ n_F(\omega - eV) - n_F(\omega ) ], 
\eeq
Differentiating $I$ with respect to $V$ gives the differential conductance
\beq
d I / dV & \propto & \int_{-\infty}^{\infty} d \epsilon \rho_S(\omega) (d n_F / d \omega) |_{\omega - eV}  \nonumber \\
\rho_S(\omega) &=& \int d k |u_k|^2  \delta(\omega - \xi_k). 
\eeq
%


\begin{thebibliography}{10}



\bibitem{ando}
S. Sasaki {\it et. al.}, arXiv:1108.1101

%
%
%
%
 \bibitem{ludwig}
 A. Schynder, S. Ryu, A. Furusaki and A. Ludwig, Phys. Rev. B \textbf{78}, 195125 (2008).
 

\bibitem{kitaev} A. Kitaev, arXiv:0901.2686 

 \bibitem{read} N. Read and D. Green, Phys. Rev. B \textbf{61}, 10267 (2000).

\bibitem{roy} R. Roy, arXiv:0803.2868 

\bibitem{zhangtsc} X. L. Qi, T. L. Hughes, S. C. Zhang, Phys. Rev.Lett. \textbf{102}, 187001 (2009) 

\bibitem{volovik} M. M. Salomaa and G. E. Volovik, Phys. Rev. B \textbf{37}, 9298 (1988); M. A. Silaev, G. E. Volovik, J. Low Temp. Phys., \textbf{161}, 460 (2010). 

\bibitem{yip} S. K. Yip, J. Low Temp. Phys., \textbf{160}, 12 (2010).

\bibitem{moore} S. Ryu, J. E. Moore and A. Ludwig, arXiv:1010.0936 

\bibitem{nagaosa} K. Nomura, S. Ryu, A. Furusaki, and N. Nagaosa, arXiv:1108.5054

\bibitem{fuberg} L. Fu and E. Berg, Phys. Rev. Lett.  \textbf{105}, 097001 (2010). 

\bibitem{qihugheszhang}  X. L. Qi, T. L. Hughes, S. C. Zhang, Phys. Rev. B \textbf{81}, 134508 (2010).

\bibitem{sato} M. Sato, Phys. Rev. B \textbf{81}, 220504(R) (2010). 

\bibitem{cava} Y. Hor {\it et al}, Phys. Rev. Lett.  \textbf{104}, 057001 (2010)

\bibitem{palee} P. A. Lee, Journal Club for Condensed Matter Physics, Feb 2010: 
http://www.condmatjournalclub.org/?p=833 

\bibitem{wray} L. A. Wray {\it et al}, Nature Physics, \textbf{6}, 855 (2010); Phys. Rev. B, \textbf{83}, 224516 (2011). 


\bibitem{heat} M. Kriener {\it et al}, Phys. Rev. Lett. \textbf{106}, 127004 (2011). 

\bibitem{sample} M. Kriener, {\it et al}, Phys. Rev. B \textbf{84}, 054513 (2011). 

\bibitem{magnetization} P. Das {\it et al}, Phys. Rev. B \textbf{83}, 220513(R) 2011). 



\bibitem{haolee} L. Hao and T. K. Lee,  Phys. Rev. B \textbf{83}, 134516 (2011)

\bibitem{abs} For reviews on surface Andreev bound states in unconventional superconductors, see S. Kashiwaya� and Y. Tanaka, Rep. Prog. Phys. \textbf{63}, 1641 (2000);  
G. Deutscher, Rev. Mod. Phys. \textbf{77}, 109 (2005). 

\bibitem{zhang} H. Zhang {\it et al}, Nature Phys. {\bf 5}, 438 (2009). 

\bibitem{liu} C. X. Liu  {\it et al}, Phys. Rev. B \textbf{82}, 045122 (2010). 

  

\bibitem{detail} See Supplemental Material at [...] for details.

\bibitem{teofukane} J. C. Y. Teo, L. Fu, and C. L. Kane. Phys Rev. B, \textbf{78}, 045426. (2008).




\bibitem{murakami} R. Takahashi and S. Murakami,  arXiv:1105.5209


\bibitem{mirror}  Strictly speaking, mirror helicity protects the second crossing along the mirror-invariant line $\Gamma M$ only. However,  
higher order terms which reduce the full rotational symmetry are small.  


\bibitem{fukane} L. Fu and C. L. Kane, Phys. Rev. B \textbf{76}, 045302 (2007).



\bibitem{anton} For a general discussion of boundary conditions for Dirac materials, see A. R. Akhmerov and C. W. J. Beenakker, Phys. Rev. B \textbf{77}, 085423 (2008).

\bibitem{qifu} Y. Qi and L. Fu, to be published. 


\bibitem{bite} J. L. Zhang {\it et al}, PNAS, \textbf{108}, 24 (2011). 

\bibitem{thallium} R. A. Hein and E. M. Swiggard, Phys. Rev. Lett. \textbf{24}, 53 (1970).

\bibitem{pbte} Y. Matsushita {\it et al}, Phys. Rev. B \textbf{74}, 134512 (2006). 

\bibitem{snte} R. Hein, Physics Letters \textbf{23}, 435 (1966).

\bibitem{gete} R. A. Hein {\it et al}, Phys. Rev. Lett. \textbf{12}, 320 (1964).

\bibitem{yptbi} N. P. Butch {\it et al}, arXiv:1109.0979

\bibitem{point} T. Kirzhner {\it et al}, arXiv:1111.5805

\bibitem{termination} Hsin Lin, private communication; Junwei Liu, private communication. 

\bibitem{greens} M. P. Lopez Sancho {\it et al}, J. Phys. F: Met. Phys. \textbf{15} (1985).

\end{thebibliography}
\end{document}